# R&D results on a CsI-TTGEM based photodetector

A. Breskin[a], V. Peskov,[b,c] A. Di Mauro[b], P. Martinengo[b], D. Mayani Paras[c],

L. Molnar[b], E. Nappi,[d] G. Paic[c], J. van Hoorne[e]

[a]Weizmann Institute of Science, Rehovot, Israel
[b]CERN, Geneva, Switzerland
[c]Instituto de Ciencias Nucleares, Universidad Nacional Autonoma de Mexico, Mexico, Mexico
[d]INFN Bari, Bari, Italy
[e]Technische Universität Wien, Vienna, Austria

**Abstract**

The very high momentum particle identification detector proposed for the ALICE upgrade is a focusing RICH using a $C_4F_{10}$ gaseous radiator. For the detection of Cherenkov photons, one of the options currently under investigation is to use a CsI coated Triple-Thick-GEM (CsI-TTGEM) with metallic or resistive electrodes. We will present results from the laboratory studies as well as preliminary results of beam tests of a RICH detector prototype consisting of a $CaF_2$ radiator coupled to a 10x10 cm$^2$ CsI-TTGEM equipped with a pad readout and GASSIPLEX-based front-end electronics. With such a prototype the detection of Cherenkov photons simultaneously with minimum ionizing particles has been achieved for the first time in a stable operation mode.

1. Introduction

The reported studies have been carried out in the framework of the ALICE Very High Momentum Particle Identification (VHMPID) RICH upgrade program aiming to extend the particle identification capability for hadrons up to 30 GeV/c. Currently, two options of possible photodetectors to be used in the VHMPID are under consideration: MWPC with a CsI photocathode (CsI-MWPC) [1] and CsI-TTGEMs [2] or its recent more robust modification: CsI-TTGEM with resistive electrodes (CsI-RETGEM)[3]. Detailed description of the TGEM and RETGEM detectors can be find in [4-6]. The main advantage of the CsI-MWPC approach is that it is based on a well proven technology, successfully exploited in the present ALICE HMPID detector. In order to minimize photoelectron losses by backscattering [7,8] and achieve high effective quantum efficiency (QE) the MWPC is operated in pure $CH_4$, which constitutes a drawback for all the safety aspects related to its flammability. In addition, in such a gas the detector stability is affected by the photon feedback which appears at gains larger than $10^4$. In contrast, CsI-TGEMs can operate in poorly-quenched gases as well as in gases which are strong UV emitters, reaching high gains without feedback problems. This also opens the possibility to use them in non-flammable gases or, if necessary, in windowless detector layouts (as in PHENIX [9]). If necessary CsI-TGEMs can operate with zero and even reversed electric field in the drift region which allows a strong suppression of the ionization signal from charged particles [10] (as also implemented in PHENIX [9]).

The aim of this work is to investigate the feasibility to exploit the hole-type multiplier option as a photodetector for the VHMPID detector and to make a preliminary comparison with the CsI-MWPC option. This investigation was performed in two stages: laboratory studies and beam tests of a small RICH prototype based on CsI-TTGEM or CsI-RETGEM.

## 2. Laboratory studies

We have recently performed a series of systematic studies of CsI-TTGEMs in the view of their potential application in UV detectors for RICH systems [11-13]. Two gas mixtures, $Ne+10\%CH_4$ and $Ne+10\%CF_4$, allowed stable operation at rather low voltages applied across each TGEM (about 500 V, close to the voltages usually applied to GEMs). In addition, the photoelectron extraction efficiency from the reflective CsI photocathode reaches about $\varepsilon_{extr}$ ~80% while the extracted photoelectron collection efficiency is $\varepsilon_{coll}$ ~ 100% [8]. The effective photon detection efficiency of the photosensitive THGEMs is:

$$E_{effph}(\lambda) = QE(\lambda)\, A_{eff}\, \varepsilon_{extr}(\lambda,V)\, \varepsilon_{coll}(V), \qquad (1)$$

where $QE(\lambda)$ is the CsI vacuum quantum efficiency as a function of the wavelength $\lambda$, $S_{eff}$ is the CsI-coated fraction of the TGEM surface and $\varepsilon_{extr}(\lambda,V)$ is the photoelectron extraction efficiency in gas. As it was shown in [12], $E_{effph}$ could reach values (at $\lambda$=170 nm) as high as 0.24 (depending on $A_{eff}$). The photon detection efficiency of CsI-TGEM could therefore reach values close to that of the CsI-MWPC presently employed in COMPASS-RICH, namely $E_{effph}\approx 0.21$ [12]. Finally, the single photoelectron detection efficiency depends on the gas gain G and on the front-end electronics threshold $q_t$:

$$E_{pe} = exp(-q_t/G) \qquad (2)$$

Therefore, for a given electronics, the detection efficiency $E_{pe}$ will increase as the detector operates at higher gains.

We have observed that the maximum achievable gain $G_{max}$ of the CsI-TGEM is not governed by the feedback loop, but by the Raether limit (see [14] for more details):

$$Q_c = G_{max} n_0 \qquad (3)$$

where $n_0$ is the number of primary electrons created in the drift volume of the detector and $Q_c$ is the critical total charge in the avalanche (typically for TGEM $Q_c \approx 10^6$ electrons). As can be seen from this formula, in the case of detection of single photoelectrons ($n_0$=1) gains up to $10^6$ are possible. However, in presence of any radioactive background, creating for example $n_0 \approx$ 10-100 electrons, the maximum achievable gain of the CsI-TGEM will be correspondingly 10-100 times smaller, respectively.

To overcome this problem the CsI-TTGEM can be operated with a reversed drift field thus achieving a gain of ~$10^5$ even under simultaneous irradiation with UV and $^{90}$Sr beta source.

For direct comparison, studies were also performed with a small MWPC featuring a geometry similar to the one used currently in ALICE and COMPASS RICH counters, with a CsI photocathode and operated in $CH_4$ at 1 atm [13]. With the reversed drift field the attainable CsI-TTGEM gain was at least 10 times larger than that of the CsI-MWPC, under similar conditions.

Very similar results were obtained in the laboratory with CsI-RETGEMs

## 3. Beam test of a small RICH prototype

Fig. 1 shows the layout of the RICH prototype tested at the CERN/PS. It consists of a $CaF_2$ window, 4 mm thick, used as Cherenkov radiator, coupled to a CsI-TTGEM or

CsI-RETGEM, flushed with Ne+10%CH$_4$ or Ne+10%CF$_4$ at p$_2$ = 1 atm. Each TGEM/RETGEM has a thickness of 0.45 mm, an active area of 10x10 cm$^2$, hole diameter of 0.4 mm (holes have 10 μm wide dielectric rims) and pitch of 0.8 mm. The readout is based on the standard ALICE HMPID electronics (GASSIPLEX and DILOGIC chips developed within the HMPID project). Event monitoring and analysis have been carried out using the ALICE software package AMORE (see for example [15]). Beam tests with an uncoated TTGEM/RETGEM and with CsI-TTGEMs /CsI-RETGEM were carried out.

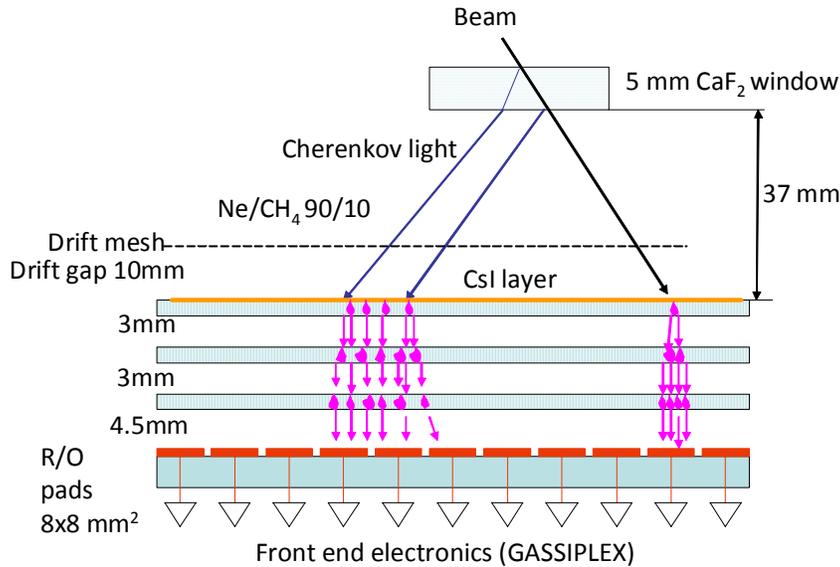

Fig1. Layout of the RICH prototype with CaF$_2$ radiator and CsI-TTGEM/CsI-RETGEM photon counter.

The TTGEM or RETGEM top electrode was coated with CsI by following the standard procedure used earlier for the ALICE HMPID photocathodes production [16]. After the evaporation of 0.3 μm thick CsI layer, the response of the TGEM/RETGEM has been evaluated by a VUV scanning system described in [16]. Fig 2 shows the mean ratio of the TGEM and RETGEM photocurrents to the reference PM as a function of the time after the evaporation. The response enhancement, few hours after the coating, related to the heat treatment, is similar to that obtained with HMPID CsI photocathodes, as well as the final photocurrent ratio [12].

Because the QE of CsI-TGEM and CsI-RETGEM were almost the same, all results obtained with these two detectors during the beam test were practically the same (under identical conditions).

After the installation into the RICH prototype, the QE of the CsI–TGEM or CsI-RETGEM has been estimated using a Hg lamp combined with a narrow band filter (185 nm) ([13,14]). The induced photocurrent has been measured between the top electrode of the first TGEM and the drift mesh as a function of the voltage applied across this gap. The saturation value of this current was then compared with the saturation value of the current produced by the same light beam in a TMAE filled detector ([17]). From these measurements the QE of the CsI photocathode was evaluated to be ~17% at 185 nm. Similar results (14-16%) were obtained in pulse counting mode (for details of these measurements see [18]).

The RICH prototype was then installed at the CERN PS/T10 beam test facility. To study the detection of minimum ionizing particles (MIP), the CsI-TGEM/Cs-RETGEM was operated with a normal drift field, reaching gains up to ~$10^5$ in a stable regime. Fig.3 shows a typical image (as was mentioned earlier, results obtained with Cs-TGEM and CsI-RETGEM were the same under identical conditions)of overlapped events obtained by a 6 GeV/c $\pi^-$ beam perpendicular to the detector operated with the normal drift field.

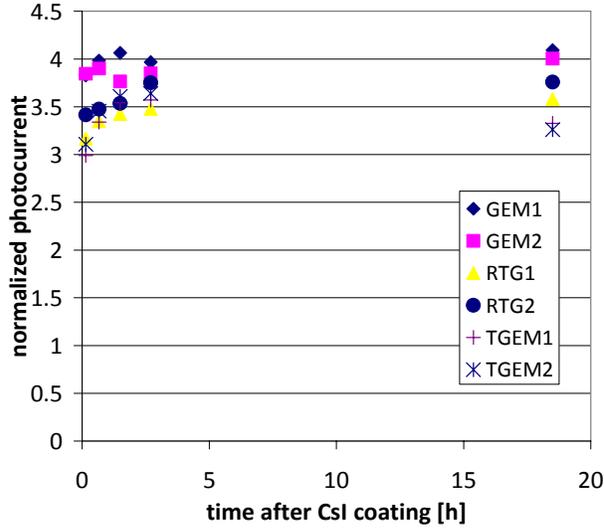

Fig. 2. Ratio of the photocurrent values measures for different CsI coated GEM, TGEM and RETGEM immediately after the evaporation and 15 hours latter( for details see[16]).

Only the MIP signal is present and no Cherenkov light is visible since it is fully trapped inside the $CaF_2$ radiator window due to total internal reflection.
Tilting the detector by an angle larger than 20° with respect to the particles direction, part of the Cherenkov cone escapes the $CaF_2$ radiator and falls in the acceptance of the top CsI-TGEM/CsI-RETGEM together with the MIP. Because of the detector configuration and the small photosensitive area, only a fraction of the produced Cherenkov radiation can be detected (Fig. 4). Fig. 5 shows single photoelectron pulse-height (PH) spectra recorded at two different HV settings, characterized by the typical exponential trend. As an example, results relative to three runs at different HV settings are summarized in Table 1. The increase in number of photon clusters is a consequence of the improvement of $E_{pe}$ which depends on the recorded gain.
 Fig 6 shows results from Monte Carlo simulations of a detector implementing a $CaF_2$ window radiator coupled to a CsI-MWPC photon detector (we used a program developed for the CsI-MWPC) operated in $CH_4$, characterized by $A_{eff}$ ~ 0.9 and $\varepsilon_{extr}$ ~ 0.9. The number of photon clusters (1.94) is larger than that observed , for example in run 2155, characterized by similar gain, due to different $A_{eff}$ (~ 0.8 for the TGEM/RETGEM) and $\varepsilon_{extr}$ (~ 0.75, in Ne+10%$CH_4$ at the electric field established by the HV setting [12]). These results confirm that the intrinsic QE of the CsI-TGEM/CsI-RETGEM is comparable with that of the ALICE present HMPID CsI photocathodes.

a)

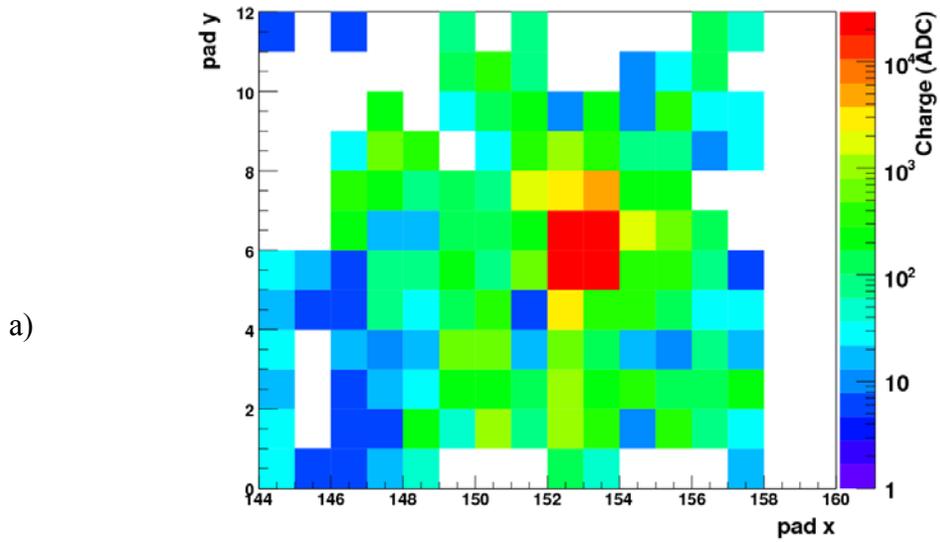

b)

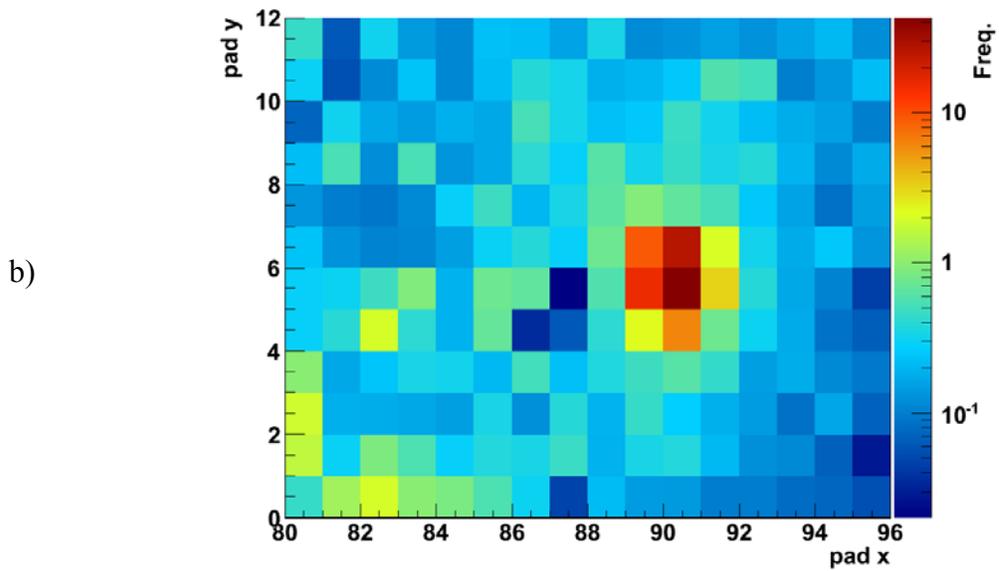

Fig. 3. Charge weighted 2D display of overlapped events when the RICH detector is perpendicular to the beam and operated with a normal drift field of -200V at a gain of $10^4$: a) CsI-TTGEM, b) CsI-RETGEM. The images of the beam is clearly seen in the central region of the screens.

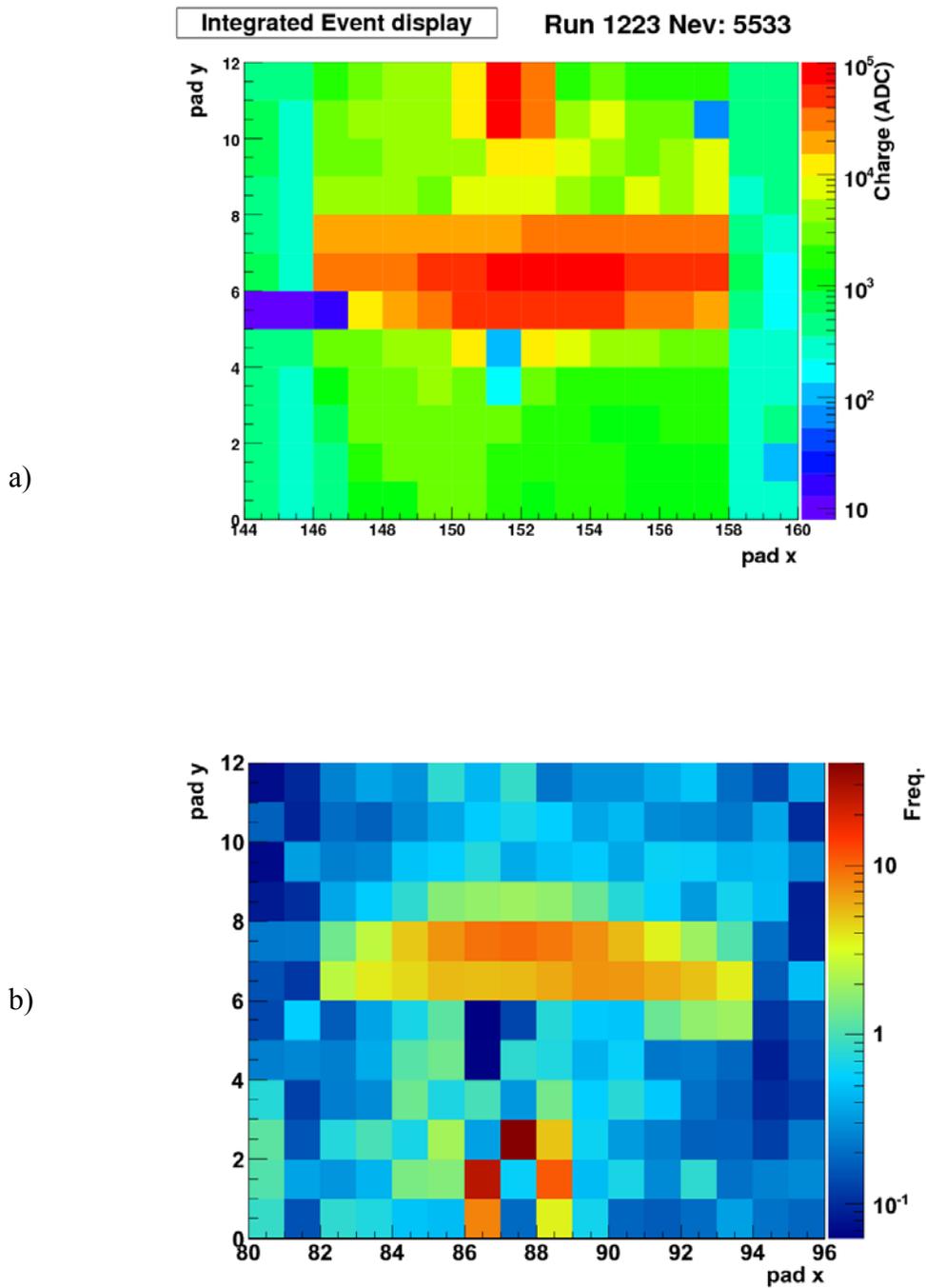

Fig. 4. A typical image of the integrated events obtained during tests of RICH prototype oriented ~32° with respect to the beam: a) CsI-TTGEM, b) CsI-RETGEM. During this particular run the CsI-TTGEM/Cs_RETGEM operated with reversed drift field (200V) at an overall gain of ~$10^5$. The spot at the top of the Fig. a) and on the bottom of Fig. b) is the image of the particle beam and the horizontal bands in the middle of each figures correspond to the detected Cherenkov photons.

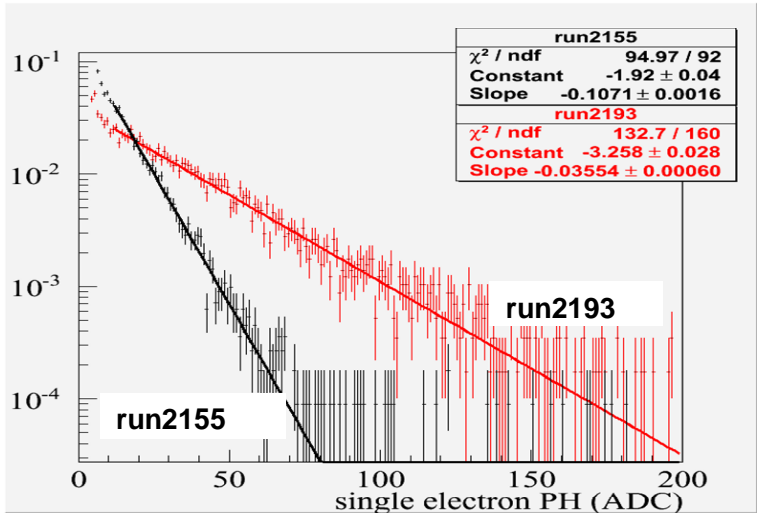

Fig. 5. Example of single electron PH distributions for two runs at different HV settings: RUN2155 - ΔV1=600, ΔV2=600, ΔV3=590; RUN 2193 - ΔV1=600, ΔV2=610, ΔV3=600 (ΔVi represent the potential difference between the bottom and top layers of the i-th TGEM).

| RUN# | Gas Mixture | Visible GAIN (ADC) | $E_{det}$ | # of photon clusters/event |
|---|---|---|---|---|
| **2155** | Ne/$CH_4$ (90/10) | 10.8 | 0.69 | 1.43 |
| **2193** | Ne/$CH_4$ (90/10) | 33.3 | 0.88 | 1.6 |
| **2201** | Ne/$CH_4$ (90/10) | 43.5 | 0.91 | 1.75 |

Table 1. Experimental conditions and main performance quantities (gain and number of photon clusters/event) for three runs shown as examples.

Preliminary results of tests with Ne+10%$CF_4$ showed a better performance in terms of photon detection efficiency, probably due to a larger extraction efficiency in this gas mixture at similar electric field at the photocathode surface [12].

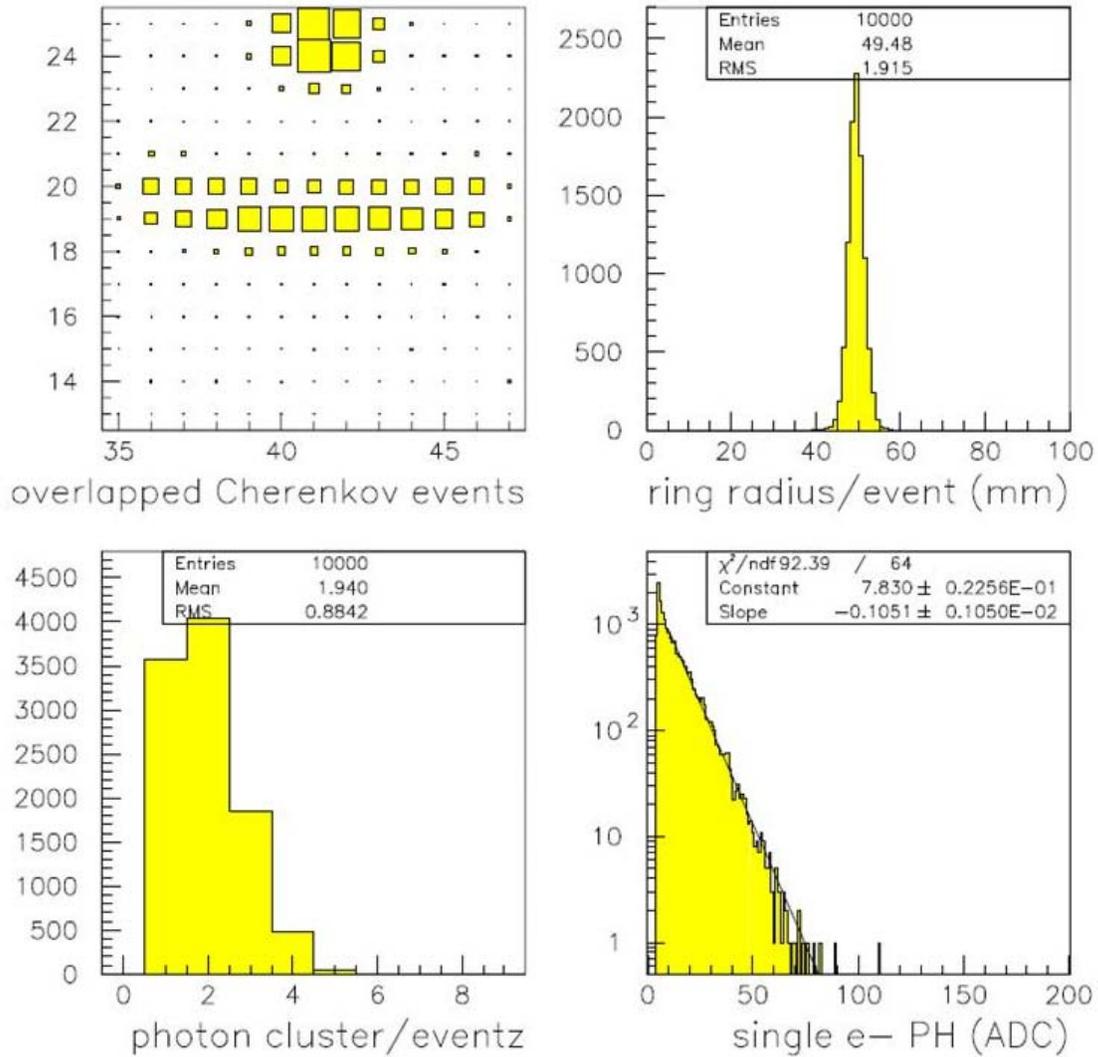

Fig. 6. Main results of simulation of RUN2155 chosen as an example. The inverse of the slope of the exponential fit of the single e- PH distribution represent the chamber gain.

### 4. Conclusions

With a CsI-TTGEM/CsI-RETGEM the detection of Cherenkov photons simultaneously with minimum ionizing particles has been achieved for the first time in a stable operation mode. The results of the performed studies clearly indicate that the CsI-TGEM/CsI-RETGEM based photodetector could be an attractive option for the ALICE VHMPID. Although the effective QE is ~30% below that of a CsI-MWPC, it can be operated at ten-fold higher gains, which under some experimental conditions, may offer higher photoelectron detection efficiency. Detailed comparison of hole-type gaseous photomultipliers with Cs-MWPC is given in [ 13] and summarized in[19].

**Acknowledgments**

G. Paic and D. Mayani Paras are acknowledging the support of the UNAM grant IN115808, the CERN-UNAM grant and the support of the Coordinacion de la Investigacion Cientifica. A. Breskin is the W.P. Reuther Professor of Research in The Peaceful Use of Atomic Energy.